\documentclass[apj,twocolumn]{emulateapj} 
\newcommand{\rayleigh}{{\sf Rayleigh}} 

\newcommand{\pr}{\ensuremath{\mathrm{Pr}}}    
\newcommand{\raf}{\ensuremath{\mathrm{Ra}_F}} 
\newcommand{\re}{\ensuremath{\mathrm{Re}}}    

\newcommand{\ptl}{\partial}
\newcommand{\DxDt}[1]{\ensuremath{\frac{D #1}{D t}}} 

\newcommand{\myvec}[1]{\ensuremath{\bm{#1}}}

\newcommand{\inner}{\myvec{\cdot}}

\newcommand{\del}{\myvec{\nabla}}
\newcommand{\myhat}[1]{\myvec{\hat{#1}}}

\newcommand{\sfrac}[2]{\mathchoice
  {\kern0em\raise.5ex\hbox{\the\scriptfont0 #1}\kern-.15em/
   \kern-.15em\lower.25ex\hbox{\the\scriptfont0 #2}}
  {\kern0em\raise.5ex\hbox{\the\scriptfont0 #1}\kern-.15em/
   \kern-.15em\lower.25ex\hbox{\the\scriptfont0 #2}}
  {\kern0em\raise.5ex\hbox{\the\scriptscriptfont0 #1}\kern-.2em/
   \kern-.15em\lower.25ex\hbox{\the\scriptscriptfont0 #2}}
  {#1\!/#2}}

\newcommand{\half}{\frac{1}{2}}

\newcommand{\overbar}[1]{\bar{#1}}

\newcommand{\grav}{\myvec{g}} 
\newcommand{\uvec}{\myvec{u}} 

\newcommand{\rhat}{\myhat{r}}      
\newcommand{\rhobar}{\overbar{\rho}} 
\newcommand{\Pbar}{\overbar{P}}
\newcommand{\Tbar}{\overbar{T}}

\newcommand{\fconv}{\ensuremath{f_{\rm conv}}} 

\usepackage{bm}
\usepackage{amsmath}
\usepackage{rotating}
\usepackage{placeins}

\usepackage{color,soul}

\usepackage{pifont}

\newcommand{\refereeEdits}[1]{#1}                           

\begin{document}

\title{Prandtl-Number Effects in High-Rayleigh-Number Spherical Convection}
\author{Ryan J. Orvedahl$^{1}$,
        Michael A. Calkins$^{2}$,
        Nicholas A. Featherstone$^{3,5}$,
        Bradley W. Hindman$^{1,4}$}
\affil{$^1$Department of Astrophysical \& Planetary Sciences, University of 
       Colorado, Boulder, CO 80309-0391, USA}
\affil{$^2$Department of Physics, University of 
       Colorado, Boulder, CO 80309-0390, USA}
\affil{$^3$Department of Applied Mathematics, University of 
       Colorado, Boulder, CO 80309-0526, USA}
\affil{$^4$JILA, University of Colorado, Boulder, CO 80309-0440, USA}
\affil{$^5$Research Computing, University of Colorado, Boulder, CO 80309-0440, USA}


\begin{abstract}
Convection is the predominant mechanism by which energy and angular momentum
are transported in the outer portion of the Sun.
The resulting overturning motions are also the primary energy source for the
solar magnetic field. An accurate solar dynamo model therefore requires
a complete description of the convective motions, but
these motions remain poorly understood. Studying stellar convection numerically
remains challenging; it occurs within a parameter regime that is extreme by
computational standards. The fluid properties of the convection zone are characterized
in part by the Prandtl number $\pr=\nu/\kappa$, where $\nu$ is the kinematic viscosity
and $\kappa$
is the thermal diffusion; in stars, \pr\ is
extremely low, $\pr\approx 10^{-7}$. The influence of \pr\
on the convective motions at the heart of the dynamo is not well understood
since most numerical studies are limited to
using $\pr\approx1$. We systematically vary \pr\ 
and the degree of thermal forcing, characterized through a Rayleigh number,
to explore its influence on the convective dynamics.
For sufficiently large thermal driving, the simulations reach a so-called convective
free-fall state where diffusion
no longer plays an important role in the interior dynamics.
Simulations with a lower \pr\
generate faster convective flows and broader ranges of scales for equivalent levels
of thermal forcing.
Characteristics of the spectral distribution of the velocity remain
largely insensitive to changes in \pr. Importantly, we find that
\pr\ plays a key role in determining when the free-fall regime is reached
by controlling the thickness of the thermal boundary layer.
\end{abstract}


\maketitle

\section{Introduction}
\label{sec:intro}

Convective motions within the outer one third of the Sun transport energy from the
radiative interior to the photosphere. In the process, these overturning
motions, which are thought to drive the solar differential rotation, play a pivotal role in
generating solar magnetic field. Any model of the solar dynamo necessarily requires a
description of the Sun's underlying convective motions, and yet those motions remain poorly
characterized in spite of the observational coverage enabled by the Sun's proximity to Earth.

On the largest spatial scales, photospheric convection manifests in celluar patterns known
as supergranules that were first noted by \citet{hart1954} and better characterized by
\citet{leighton1962}.
The horizontal extent of these cells is approximately 35 Mm, and they possess a
spectral peak in photospheric Dopplergram power around spherical harmonic degree
$\ell\approx 120$
\citep[e.g.,][]{hathaway2000, hathaway2015}. In addition, smaller-scale motions known as
granulation are clearly visible in the photosphere, possessing a characteristic size of
about 1 Mm and a clear peak in photospheric Dopplergram power around
$\ell\approx 10^3$ \citep{bray1984}.

Presently, only the granular component of photospheric convection is reliably captured in
direct numerical simulations. Radiative hydrodynamic simulations of solar surface
convection that can simulate granulation fail to yield clear evidence for supergranulation
\citep[e.g.,][]{stein2009, ustyugov2010}.
Inconsistencies between numerical models and the Sun have also been observed in the velocity
power distribution associated with larger scales of convection
\citep[e.g.,][]{miesch2008}. Measurements of deep
convective flow speeds, made using time-distance helioseismology, suggest that convection
models may overestimate the amplitude of the convection on spatial scales larger than 30
Mm \citep{hanasoge2012}. The results of \citet{hanasoge2012}
estimate that the convective velocities on spatial scales larger than 70 Mm is at most
5-6 m s$^{-1}$, about an order of magnitude weaker than that expected from simulations or
theoretical arguments \citep[e.g.,][]{miesch2012}. Ring-analysis measurements of the
subsurface flows in the near-surface shear
layer, however, exhibit good agreement with models and theory \citep{greer2015}. 

Resolving these descrepencies requires careful comparison of these two different
helioseismic techniques, and perhaps improvements to both. A resolution to this problem
also requires a better theoretical understanding of convective dynamics under stellar
conditions. Exploring stellar convection numerically remains challenging owing to
the fact that it occurs within parameter regimes that are considered extreme by modern
computational and laboratory standards alike. These regimes can be characterized by several
nondimensional parameters. In particular, the Reynolds number measures the relative
importance of inertial forces to viscous forces, the Rayleigh number expresses the relative
strength of buoyancy driving and diffusive effects, and the Prandtl number specifies the
relative importance of viscosity to thermal diffusion.
Estimates for the Sun lead to
values of the Reynolds, Rayleigh, and Prandtl numbers on the order of $10^{13}$, $10^{20}$,
and $10^{-7}$, respectively, indicating the solar convection zone is highly turbulent
\citep[e.g.,][]{ossendrijver2003}. 

Such extreme values of nondimensional parameters are largely an expression of the fact that
while diffusion in stellar interiors may be active on very small scales, it tends to be
negligible at the system scale. Achieving such a situation, wherein diffusion plays no
appreciable role in the leading-order force balance, is possible in those parameter regimes
already accessible through computational models. \citet{featherstone2016, featherstone2016b}
identified two such regimes by exploring the response of convection to changes in Rayleigh
number
and
Ekman number (which expresses the relative importance of the Coriolis force and viscous
diffusion), while using a fixed Prandtl number of unity.
In so doing, they identified asymptotic scaling laws for spectral properties of the
convection that could in principle be extrapolated to the stellar parameter regime.
The purpose of this paper is to extend those studies by exploring the response of the
convective spectrum to changes in the Prandtl number.

This paper is organized as follows. In Section \ref{sec:prandlt_disc}, we summarize
earlier
investigations into the role of the Prandtl number. In Sections \ref{sec:num_model} and
\ref{sec:num_experiment}, we discuss our numerical model and the parameter space
that was explored. Results are discussed in Section \ref{sec:results},
followed by a discusion of their implications in Section \ref{sec:perspective_conclusions}.

\subsection{Prandtl Number in Convection}
\label{sec:prandlt_disc}
The Prandtl number $\pr=\nu/\kappa$, where $\nu$ is the kinematic viscosity and $\kappa$
is the thermal diffusivity, is known to be small in the dynamo regions of
planets and stars \citep{ossendrijver2003,roberts2007}. It is well known from linear theory
that the value of the Prandtl number both influences the structure and amplitude of
convective motions and controls the critical Rayleigh number required for the onset of
rotating convection \citep[e.g.,][]{chandrasekhar1961}. Asymptotic approximations
to the governing
fluid equations can be carried out based on the size of the Prandtl number. Such
approximations yield some insight into the convective dynamics arising under different
Prandtl number regimes. In the large Prandtl number limit, the influence of inertia is weak
and can be neglected; this limit is routinely exploited in studying the convection
of planetary mantles where the Prandtl number can be
of order $10^{20}$ \citep{schubert2001}.
\citet{spiegel1962} developed
an approximate set of equations valid in the limit $\pr \rightarrow 0$ that showed
inertia plays a leading-order role in the convective dynamics
\citep[see also][]{thual1992}. Both of these approximate models have been applied only to
nonrotating and incompressible Boussinesq systems to date. 


Much less is known about the role of the Prandtl number in compressible convection. Earlier
studies of compressible convection have primarily used Prandtl number of order unity
\citep[e.g.,][]{gilman1977,gilman1981,goudard2008,christensen2011,soderlund2012,
schrinner2012,gastine2015,gastine2016,wicht2016}. While there are also many studies that
make use of nonunity Prandtl numbers
\citep[e.g.,][]{brown2011,kapyla2013,nelson2014,jones2014,
augustson2015,augustson2016,duarte2016,brun2017}, no parameter studies that vary the Prandtl
number systematically have been carried out. One exception to this trend is the work of
\citet{omara2016}, who explored the characteristics of high Prandtl
compressible convection. These authors found that high Prandtl convection tended to possess
lower characteristic flow-speeds with respect to
unity Prandtl number convection, owing to the enhanced entropy
content of its downflow plumes. Finally, we note that recent work using a small Prandtl
number with rapid rotation has found that the anelastic approximation can yield spurious
behavior \citep{calkins2015} that does not appear in nonrotating anelastic convection
\citep{calkins2015b}. These results raise serious questions regarding the applicability of
the anelastic approximation within rotating stellar interiors where it remains to be seen
if the convective flows are well-approximated by a Prandtl number of unity. 

Systematic parameter space studies of convective dynamics in stellar interiors have, so far,
focused largely on the role of buoyancy driving and rotation.
\citet{featherstone2016} investigated the response of the convection to varying
Rayleigh number and varying degrees of density stratification. Those simulations
were nonrotating,
hydrodynamic, $\pr=1$ and demonstrated a clear scaling relationship between
kinetic energy and Rayleigh number. Those results also suggest that a naive interpretation
of model results (by ascribing solar values to all problem parameters but the diffusion
coefficients) will naturally overestimate the low-wavenumber power in the convective power
spectrum.
The influence of rotation was investigated using a similar methology by
\citet{featherstone2016b} who identified a complementary scaling law relating
convective-cell-size and rotational influence. When rotation is present, and diffusive
effects are negligible, the typical spatial-size of convective cells is determined primarily
through the Rossby number, which expresses the ratio of the rotation period to a
characteristic convective time scale. Their work was also restricted to $\pr=1$.

Through this paper, we extend these studies and examine the effects of Prandtl number
variation on the convective dynamics. We present a series of numerical
simulations designed to examine how the structure and amplitude of the convective flow
within a stellar interior depends on the Prandtl number and the convective forcing.
We vary the Prandtl number and the convective forcing in a systematic way, covering both
low and high Prandtl numbers. Effects due to rotation and magnetism are not included.
We will show that the convection develops smaller-scale structures as the convective forcing
is increased and the Prandtl number is decreased, corresponding to an increase in
high-wavenumber power. As the high-wavenumber power increases, the low-wavenumber power
decreases and this trend occurs for all Prandtl numbers studied. We also show that the
Prandtl number has an important influence on the boundary-layer thickness.

\section{Numerical Model}
\label{sec:num_model}
This study is based on a series of 3D, nonlinear convection simulations that use the
pseudo-spectral convection code \rayleigh\ \citep[e.g.,][]{featherstone2016}. We employ a
spherical
geometry and represent the horizontal variation of all variables along spherical
surfaces using spherical harmonics $Y_\ell^m(\theta,\phi)$. Here $\ell$ is the
spherical harmonic degree, and $m$ is the azimuthal mode order. In the radial direction,
we employ a Chebyshev collocation method, expanding all variables in
Chebyshev polynomials $T_n(r)$, where $n$ is the degree of the polynomial.

We are particularly interested in understanding convection in the deep stellar interior,
far removed from the photosphere. In this region, plasma motions are subsonic and
perturbations to thermodynamic variables are small compared to their mean,
horizontally averaged values (represented using overbars). Under these conditions,
the anelastic approximation provides a convenient description of the system's
thermodynamics \citep{gough1969,gilman1981}. The governing evolution equations
include the continuity equation
\begin{equation}
 \del\inner\left(\rhobar\uvec\right) = 0,
\end{equation}
where $\rhobar$ is the background density and $\uvec$ is the fluid velocity.
The momentum equation is given by
\begin{equation}
 \DxDt{\uvec} = -\del\frac{P}{\rhobar} - \frac{S}{c_p}\grav
                + \frac{1}{\rhobar}\del\inner\mathcal{D},
\end{equation}
where $P$ is the pressure, $S$ is the entropy,
$c_p$ is specific heat at constant pressure,
$\grav$ is the gravitational acceleration,
and the viscous stress tensor $\mathcal{D}$ is given by
\begin{equation}
 \mathcal{D}_{ij} = 2\rhobar\nu\left[e_{ij} - \frac{1}{3}\ptl_ku_k\delta_{ij}\right].
\end{equation}
Here, $e_{ij}$ is the strain rate tensor and $\delta_{ij}$
is the Kronecker delta. Written in terms of the entropy, the thermal energy equation
is given by
\begin{align}
 \rhobar\Tbar\DxDt{S} = &\del\inner\left(\rhobar\Tbar\kappa\del S\right) + Q\left(r,\theta,\phi\right) \\
        &+ 2\rhobar\nu\left[ e_{ij}e_{ij} - \frac{1}{3}\left(\del\inner\uvec\right)^2\right],
\end{align}
where $\Tbar$ is the background temperature.
Sources of internal heating and cooling are encapsulated in the functional form of $Q$.
A linearized equation of state closes the system and is given by
\begin{equation}
 \frac{\rho}{\rhobar} = \frac{P}{\Pbar} - \frac{T}{\Tbar}
                      = \frac{P}{\gamma\Pbar} - \frac{S}{c_p},
\end{equation}
assuming the ideal gas law
\begin{equation}
 \Pbar = \mathcal{R}\rhobar\Tbar,
\end{equation}
where $\mathcal{R}$ is the gas constant and $\gamma=5/3$ is the adiabatic index.

\section{Numerical Experiment}
\label{sec:num_experiment}

We have constructed a set of 34 model stellar convection zones designed to explore
how the convective kinetic energy depends on both the thermal diffusion and the
viscous diffusion as characterized by the Prandtl number. The diffusion coefficients are
taken to be constant
values within each simulation. In particular, they have no variation with radius.
Table 1 has a detailed list of all model parameters for each run.

Each model is constructed using a polytropic background state following \citet{jones2011}.
The background states were constructed in a similar fashion to the models presented
in \citet{featherstone2016}. We use a polytropic index of $n=1.5$,
which corresponds to the adiabatic value, and model the innermost
3 density scale heights of the convection zone. The spherical shell has an aspect ratio
of $\chi = r_i/r_o = 0.759$ corresponding to a dimensional shell depth of 159 Mm,
where $r_i$ and $r_o$ are the inner and outer radii of the domain, respectively.

Our models are fully characterized by two parameters: a Rayleigh number and a Prandtl number.
As discussed in \citet{featherstone2016}, a flux Rayleigh number \raf\ appropriate for this
system may be defined as
\begin{equation}
 \label{eqn:Ra}
 \raf = \frac{\tilde{g}\tilde{F}H^4}{c_p\tilde{\rho}\tilde{T}\nu\kappa^2},
\end{equation}
where tildes indicate volume averages over the full shell, making \raf\ a bulk Rayleigh
number. In this definition, $F$ is the thermal energy imposed by the radiative heating
and $H$ is chosen to be the shell depth, $r_o-r_i$.
The nondimensionalization was
carried out using the shell depth and the viscous diffusion timescale $H^2/\nu$.

Heat enters the system through the internal deposition by $Q$, which drops to zero at
the upper boundary. In all simulations we adopt a functional form of $Q$ that depends only
on the background pressure profile such that
\begin{equation}
Q\left(r,\theta,\phi\right) = \alpha\left(\Pbar\left(r\right) - \Pbar\left(r_o\right)\right).
\end{equation}
The normalization is defined so that
\begin{equation}
L_\star = 4\pi\int_{r_i}^{r_o} Q\left(r\right) r^2\ \mathrm{d}r,
\end{equation}
where $L_\star$ is the stellar luminosity. The thermal energy flux $F\left(r\right)$ that
convection and conduction must transport across a spherical surface at radius $r$ is then
given by
\begin{equation}
F\left(r\right) = \frac{1}{r^2}\int_{r_i}^{r} Q\left(x\right) x^2\ \mathrm{d}x.
\end{equation}

For all simulations, we have adopted impenetrable and stress-free boundary conditions
on the velocity. The radial entropy gradient is forced to vanish at the lower boundary
of the convection zone, and the entropy perturbations are forced to vanish at
the upper boundary.

Our numerical experiments span the range of
$4\times 10^2 \lesssim \raf \lesssim 7\times 10^6$ and $0.1 \le \pr \le 4$.
Each simulation was initialized using a
small random thermal perturbation, evolved until the kinetic energy reached a
statistically steady state, and further evolved for at least one diffusion time. Since
there are two diffusion time scales, the larger of the two was used for this purpose.
The larger of the two diffusion times is also the time interval over which a time
average is computed when necessary.
This averaging interval includes several tens of convective overturning times.

\begin{figure*}[ht]
{\centering
\centerline{\includegraphics[scale=0.6,angle=0]{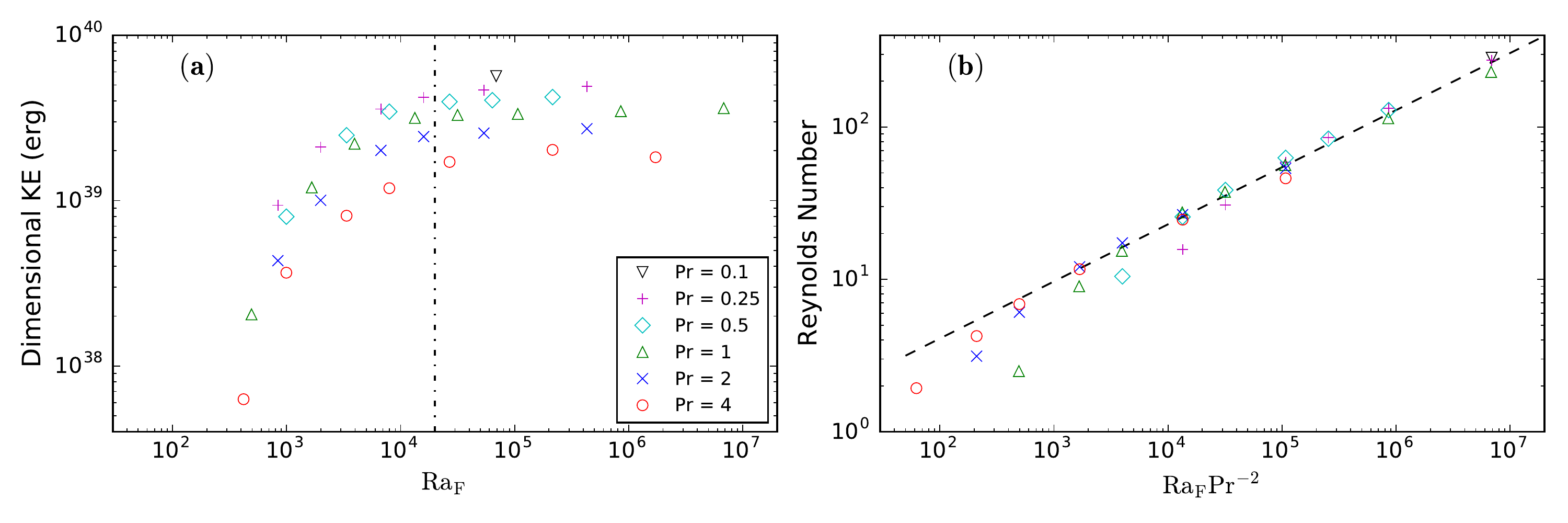}}
\caption{(a) Dimensional kinetic energy vs. flux Rayleigh number
$\raf$ for all cases. Colored symbols indicate different Prandtl numbers.
Low Prandtl number runs have higher kinetic energies for a given \raf.
Beyond the high-\raf\ cutoff, denoted by the vertical dot-dashed line,
the kinetic energy tends toward an asymptotic value.
(b) Reynolds number vs. $\raf\pr^{-2}$. The high-\raf\ region reaches an
asymptotic regime with a power law scaling exponent that is very close to $1/3$.
The dashed line is $\left(\raf\pr^{-2}\right)^{0.373}$.
Colored symbols are the same as in panel (a).
}
\label{fig:dimKE}
}
\end{figure*}
\section{Survey of Results}
\label{sec:results}
\subsection{Kinetic Energy Scaling}
\label{sec:KE_scale}
We begin our examination of the convective energetics by looking at the integrated
kinetic energy KE, defined as
\begin{equation}
 \label{eqn:KE}
 \mathrm{KE} = \frac{1}{2}\int \rhobar\left(r\right)
           \left|\uvec\left(r,\theta,\phi\right)\right|^2\ \mathrm{d}^3x,
\end{equation}
where the integration is computed over the entire domain and then time averaged.

Figure \ref{fig:dimKE}(a) shows the dimensional kinetic energy vs. \raf. Different
symbols indicate different values of \pr. As the Prandtl number is lowered, the
kinetic energy increases for any given \raf. The kinetic energy for those runs with
$\raf\gtrsim 2\times 10^4$ appears to have reached a steady value that is independent of \raf.
The level at which the kinetic energy saturates is dependent on the Prandtl number.
The saturation of the kinetic energy as the Rayleigh number is changed was also
found in \citet{featherstone2016}, although their study was restricted to a Prandtl number of
unity.

As \pr\ is increased, $\nu$ becomes larger leading to enhanced viscous dissipation and
a smaller Reynolds number. The Reynolds number measures the relative ratio of inertial
forces to viscous forces and is given by
\begin{equation}
 \re = \frac{\sqrt{\widetilde{\left|\uvec\right|^2}} H}{\nu},
\end{equation}
where $H$ and the tilde retain the same meaning as before, representing the shell depth
and a volume average, respectively. Larger Prandtl numbers will produce smaller Reynolds
numbers for a given \raf.

Figure \ref{fig:dimKE}(a) indicates that below some \raf\ cutoff, diffusion plays
a leading-order role in the force
balance. Beyond $\raf\sim 2\times 10^4$, which we denote the ``high-\raf'' region\footnote{
This is the same cutoff used in \citet{featherstone2016}. Their reported \raf\ are too high
by a factor of $\pi$; we correct for that here.}, this is no longer the case;
diffusion no longer plays a leading-order role in the global force balance and the
kinetic energy remains constant as \raf\ is increased. We note that the cutoff for the
high-\raf\ region is based on the $\pr=1$ results of \citet{featherstone2016}.
Importantly, the cutoff is a decreasing function of the Prandtl number; lower Prandtl
numbers will have a lower high-\raf\ cutoff. To simplify the analysis, only a
single cutoff is used.

The kinetic energy can also be discussed from a nondimensional point of view. To do this,
we choose a nondimensional measure of the kinetic energy $\widehat{\mathrm{KE}}$ as
\begin{equation}
 \label{eqn:nondimKE}
\widehat{\mathrm{KE}} \equiv \frac{H^2}{\nu^2 M}\mathrm{KE}.
\end{equation}
The nondimensionalization has been carried out using the mass $M$ contained within
the spherical shell, the shell depth, and the viscous diffusion timescale. Under
such a nondimensionalization, the kinetic energy can be related to the Reynolds
number as
\begin{equation}
\widehat{\mathrm{KE}} \sim \re^2.
\end{equation}
Therefore we do not plot the nondimensional kinetic energy separately, but do list it
for each run in Table 1.

If \raf\ is large enough to be in the high-\raf\ regime, then the kinetic energy is
largely insensitive to the level of diffusion. This implies that the velocity is not
strongly dependent on the level of diffusion and that the Reynolds number should scale
with the viscosity as $\re\sim \nu^{-1}$. Given our definition of \raf\ and \pr,
namely that $\raf\sim\nu^{-1}\kappa^{-2}$ and $\pr\sim\nu\kappa^{-1}$, the Reynolds
number scaling becomes $\re\sim\raf^{1/3}\pr^{-2/3}$.
In Figure \ref{fig:dimKE}(b), we plot the Reynolds number vs. $\raf\pr^{-2}$,
where the different symbols are the same as in panel (a). Each data point
is time averaged. A least-squares fit to the data
in the high-\raf\ region yields a scaling law of
\begin{equation}
 \re \propto \left(\raf\pr^{-2}\right)^{0.373\pm 0.008}.
\end{equation}

\begin{figure*}[ht]
{\centering
\centerline{\includegraphics[scale=0.8,angle=0]{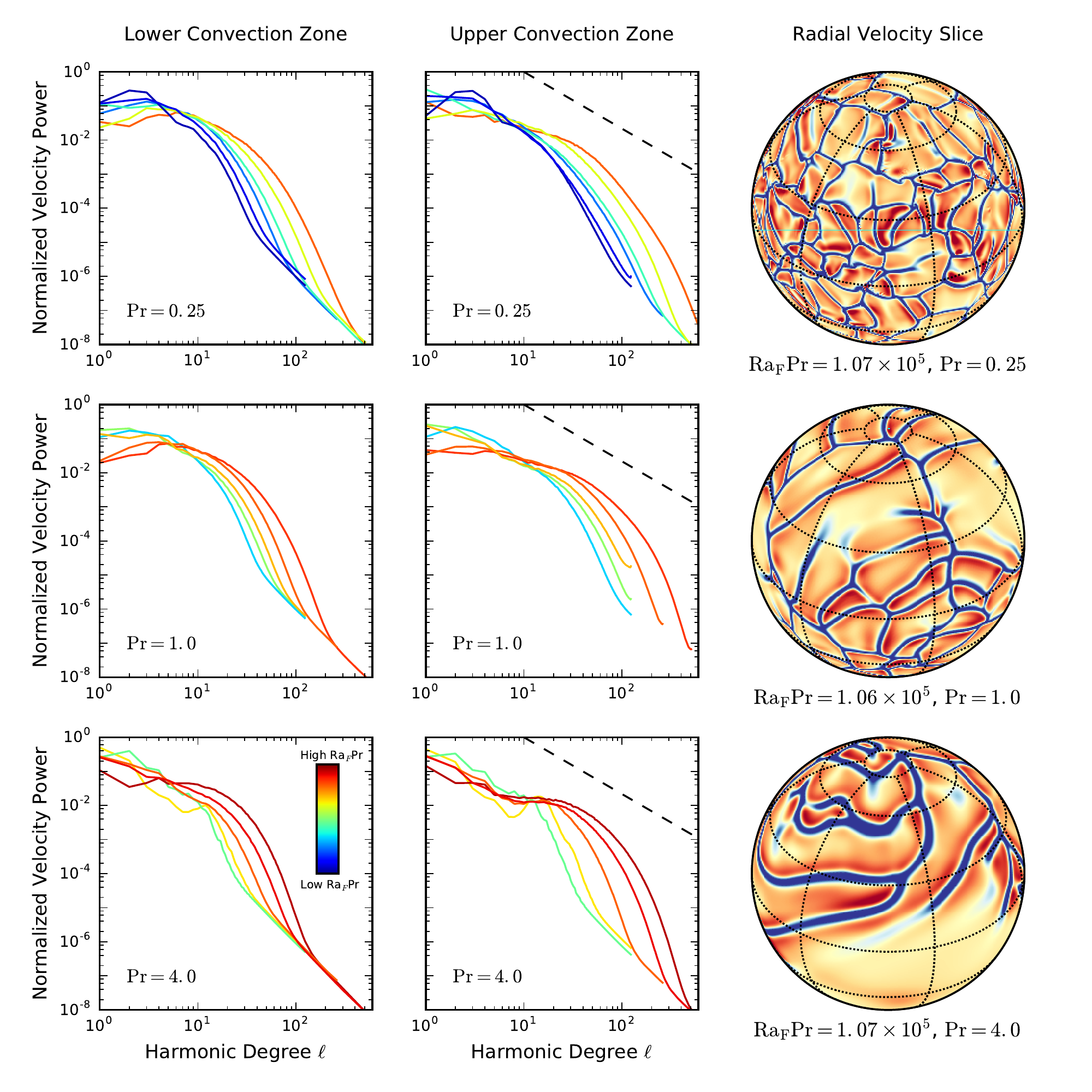}}
\caption{Time-averaged velocity power spectra and shell slices
for cases with $\pr=0.25$, $\pr=1.0$, and $\pr=4.0$. The power spectra have been
normalized such that each curve has unit integrated power. Each row
represents a different value of \pr. The first two columns correspond to power spectra
at a single depth within the convective shell. The dashed black line in the
center column has a slope of $-5/3$ for reference. Within each panel, spectra
for all cases at that depth and \pr\ are displayed. Each curve is colored by $\raf\pr$
with low $\raf\pr$ in blue tones and high $\raf$ cases are in red tones. As $\raf$ is
increased, the power at low $\ell$-values increases initially. At high $\raf$,
it decreases as high-wavenumber power is generated at the expense of low-wavenumber
power. This trend occurs at all Prandtl numbers studied. The last column shows shell
slices of the radial velocity taken near the outer boundary at $r/r_o\approx 0.985$.
The larger \pr\ run shows wider down flow lanes compared to the small \pr\ run.}
\label{fig:power_spectra}
}
\end{figure*}
These results indicate that there are two distinct parameter regions; one in which
diffusion is an important factor in the global force balance (the low-\raf\ region),
and one where diffusion no longer plays an appreciable role in the interior,
bulk global force balance (the high-\raf\ region).

\subsection{Spectral Distribution}
We find the kinetic energy may reach a \raf-independent regime, but the flow's morphology
is still affected by the level of diffusion. This can be seen in the relative
spectral distribution of velocity between the high- and low-\raf\ systems.
Figure \ref{fig:power_spectra} shows
the velocity power spectra for all runs with Prandtl numbers of $\pr=0.25$, $\pr=1.0$,
and $\pr=4.0$. Each spectrum has been normalized such that it has unit
integrated power.
The rows correspond to the different Prandtl numbers. The first column
shows the spectra taken at the lower convection zone or $r/r_o \approx 0.775$.
The second column shows the spectra near the upper boundary, or $r/r_o \approx 0.985$,
in the thermal boundary layer.
In each panel, all spectra with the given Prandtl number are plotted.
Each spectrum is colored by $\raf\pr$, with high values taking on red tones and
low values displaying blue tones. Each spectrum is a time average over several tens
of overturnings.

\begin{figure*}[ht]
{\centering
\centerline{\includegraphics[scale=0.6,angle=0]{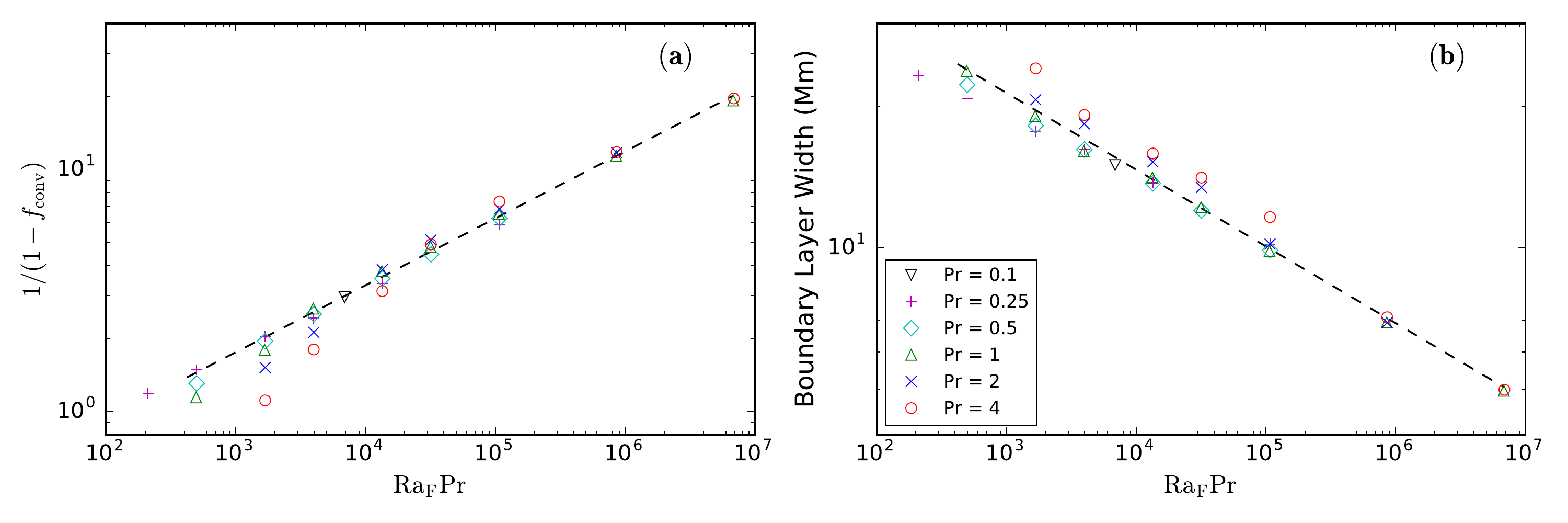}}
\caption{Energy transport and thermal boundary layer scaling with \raf\pr.
(a) Fractional convective flux $1/\left(1-\fconv\right)$ vs. \raf\pr. The dashed line is
$\left(\raf\pr\right)^{0.275}$, very close to a 2/7 scaling. (b) Thermal boundary layer
width plotted vs. \raf\pr. The dashed line is
$\left(\raf\pr\right)^{-0.164}$, very close to -1/6.}
\label{fig:fconv}
}
\end{figure*}
Across all cases, as \raf\ increases from low values, the point-wise velocity power
increases at nearly all spherical harmonic degrees. The $\ell$ value associated
with the peak for each spectra increases with increasing \raf. This indicates smaller
scale structures become more apparent with increasing \raf.
At sufficiently high-Rayleigh-number, power in the high-$\ell$ portion
continues to increase, but the low-$\ell$ portion starts to decrease. This suggests
a break down of large-scale coherent structures. The highest
Rayleigh number runs have lower power in the large scales compared to the low-\raf\
counterparts. The ``low-wavenumber'' region occurs in the approximate range of
$\ell\lesssim 10$ for most simulations in this study. This trend occurs for all
values of \pr\ that were studied. If viscosity played a significant role in the
asymptotic regime, one might expect there to be some variation in the spectra when
the Prandtl number is varied. We do not observe large variations between the spectra
for the Prandtl numbers within our range indicating that viscosity only plays a
minor role in the asymptotic regime.

The third column in Figure \ref{fig:power_spectra} plots shell slices of the radial
velocity taken near the upper boundary, the same location in radius as the second column.
Each slice is a single snapshot in time.
The color scale is the same for all three shell slices with red tones indicating positive,
outward flows and blue tones indicating negative, inward flows.
The lower Prandtl number run (the top panel) displays larger velocities
and more small-scale structures compared
to the larger Prandtl cases. The Reynolds numbers of all three slices cover a large range.
The top row $\pr=0.25$ run has $\re=261.2$, the middle slice with $\pr=1.0$
has $\re=56.3$, and the bottom slice with $\pr=4.0$ has $\re=11.7$. The large range in
Reynolds numbers indicates that the inertial subrange for each simulation is different.

At sufficiently high Rayleigh number, the integrated KE becomes independent of the level
of diffusion (both thermal and momentum diffusion). As $\nu$ is decreased and \raf\ is
increased, the flow becomes more turbulent with smaller scale structures.
To leading-order, once the high-\raf\ regime is reached, the total integrated
dimensional KE is constant as the diffusion is further reduced.
This fact is largely independent of $\pr$ for the
Prandtl numbers that were within our range, however, a weak \pr\ dependence
remains because the inertial subrange is extended as the viscosity is reduced individually.
These trends are similar to those found in the simultions of \citet{featherstone2016},
which used a Prandtl number of unity.

\subsection{Energy Transport}
\label{sec:E_transport}
The energy transport across the layer can be characterized by four radial energy
fluxes: the enthalpy flux $F_e$, the kinetic energy flux $F_\mathrm{KE}$, the conductive
flux $F_c$, and the viscous flux $F_\nu$, which we define as
\begin{equation}
 F_e = \rhobar c_p u_r T,
\end{equation}
\begin{equation}
 F_\mathrm{KE} = \half\rhobar u_r \left|\uvec\right|^2,
\end{equation}
\begin{equation}
 F_c = \kappa \rhobar \Tbar \frac{\ptl S}{\ptl r},
\end{equation}
\begin{equation}
 F_\nu = -\left(\uvec \inner \mathcal{D} \right)\inner\rhat,
\end{equation}
respectively. Note that the conductive flux $F_c$ is associated with the diffusion of
entropy perturbations, and it should not be confused with radiative diffusion arising
from the reference state temperature gradient; that effect is represented by $Q$ in our
models. Averages are taken of these fluxes over several diffusion times, indicated
using brackets. We consider the contribution of conduction by looking at the 
fractional convective flux $\fconv$, defined as
\begin{equation}
 \label{eqn:fconv}
 \fconv \equiv
 \frac{\int \left<F_e + F_\nu + F_\mathrm{KE}\right>\ \mathrm{d}^3x}
      {\int \left<F_c + F_e + F_\nu + F_\mathrm{KE}\right>\ \mathrm{d}^3x}
\end{equation}
Upon rearranging this quantity, we can write
\begin{align}
 \frac{1}{1-\fconv} &=
 \frac{\int \left<F_c + F_e + F_\nu + F_\mathrm{KE}\right>\ \mathrm{d}^3x}
      {\int \left<F_c\right>\ \mathrm{d}^3x} \\
 &=
 1 + \frac{\int \left<F_e + F_\nu + F_\mathrm{KE}\right>\ \mathrm{d}^3x}
          {\int \left<F_c\right>\ \mathrm{d}^3x}
\end{align}
\begin{figure}[ht]
{\centering
\centerline{\includegraphics[scale=0.7,angle=0]{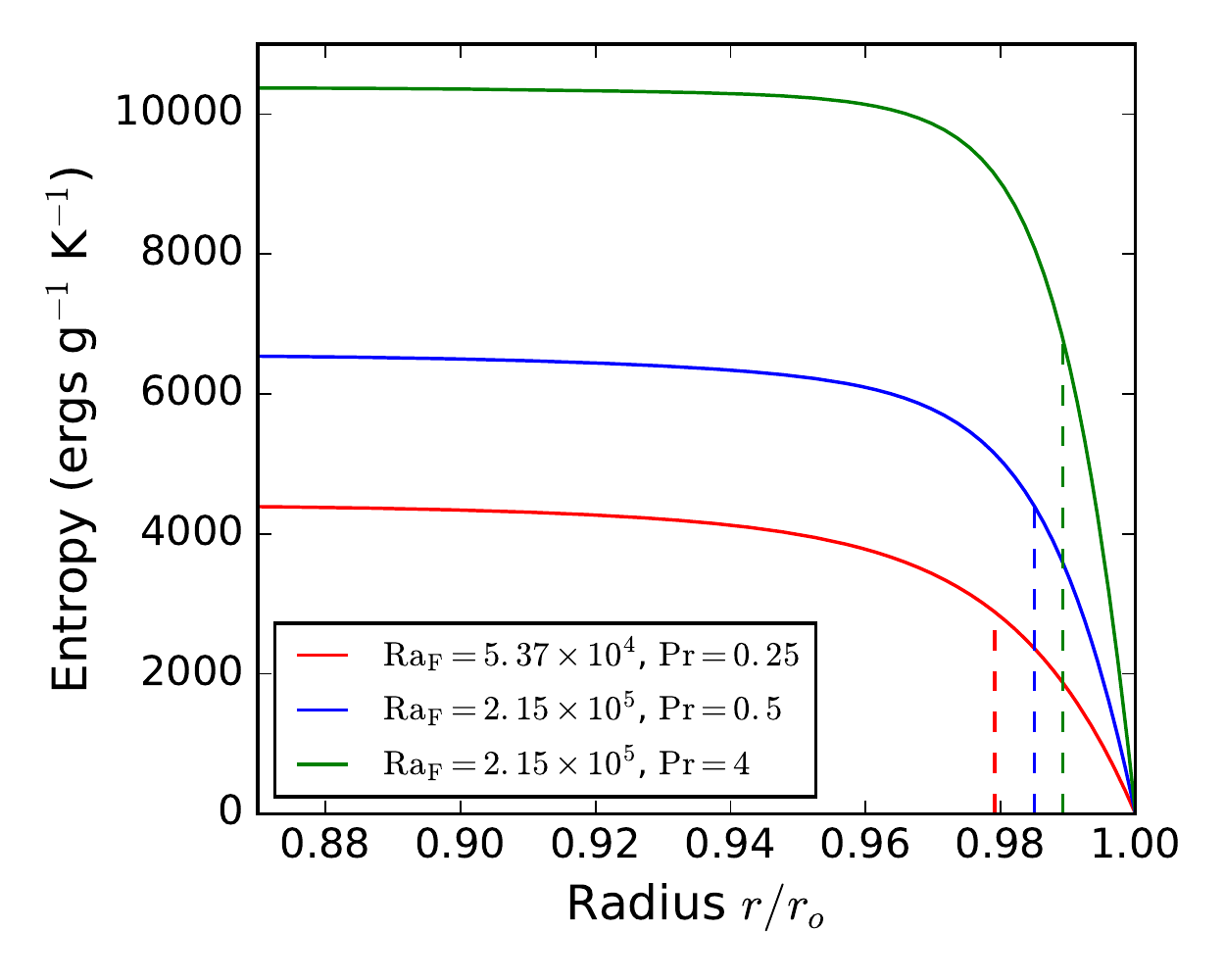}}
\caption{Time-averaged mean entropy profiles for three cases in the
high-\raf\ regime. The vertical dashed lines indicate the location of the boundary layer as
calculated using Equation \ref{eqn:wBL}.}
\label{fig:S_wBL}
}
\end{figure}
This resembles the traditional Nusselt number, but differs in two important ways.
Firstly, the traditional Nusselt number does not include the viscous flux or the kinetic
energy flux, both of which we include. Secondly, the conductive flux that appears in our
definition is the \emph{established} conductive flux, not the conductive flux in the
absence of convection.
Values of $1/\left(1-\fconv\right)$ that are of order unity translate to a lack of
convective heat transport.
Large values indicate convection plays a dominant role over thermal conduction in
transporting energy through the shell.
Figure \ref{fig:fconv}(a) plots $1/\left(1-\fconv\right)$ as a function of $\raf\pr$.
Viscosity is not expected to play a large role in the heat transport across the shell,
which is why we plot against $\raf\pr$ and not simply $\raf$.
The plot shows a clear trend that can be fit using least-squares to obtain
\begin{equation}
 \frac{1}{1-\fconv} \propto \left(\raf\pr\right)^{0.275\pm 0.002}
\end{equation}
with a scaling exponent that is approximately $2/7$ ($\sim 0.286$).
Other studies have found that the
Nusselt number scales with the Rayleigh number to the $2/9$ power ($\sim 0.222$)
\citep[e.g.,][]{gastine2015}, when a flux based Rayleigh number is used.
Our results have a steeper exponent because we include the viscous flux and
the kinetic energy flux, which act to increase $1/\left(1-\fconv\right)$
at any given Rayleigh number.

\subsection{Boundary Layer Thickness}
We can further characterize the role of conduction by determining how
\raf\ and \pr\ control the thermal boundary layer thickness. We define the thermal
boundary layer thickness in terms of the time-averaged mean entropy, such that
\begin{equation}
 \label{eqn:wBL}
\delta_\mathrm{BL} = \int \frac{\mathrm{sup}\left<S\left(r\right)\right>
                                     - \left<S\left(r\right)\right>}
                          {\mathrm{sup}\left<S\left(r\right)\right>}
                \ \mathrm{d}r,
\end{equation}
where the brackets indicate a time-average as before and $\mathrm{sup}$ indicates the
supremum. Figure \ref{fig:S_wBL} shows three examples of the time-averaged mean
entropy profiles and the associated boundary layer location as calculated using Equation
\ref{eqn:wBL}.
We plot the variation of $\delta_\mathrm{BL}$ with \raf\pr\ for different Prandtl numbers
in Figure \ref{fig:fconv}(b). We plot against \raf\pr\ because we do not expect the
thermal boundary to depend on the viscosity and the quantity \raf\pr\ is independent
of the viscosity and scales as $\raf\pr\sim\kappa^{-3}$.
A least-squares fit gives the scaling law
\begin{equation}
 \label{eqn:wBLfit}
 \delta_\mathrm{BL} \propto \left(\raf\pr\right)^{-0.164\pm 0.003}.
\end{equation}
The scaling exponent is very close to -1/6 ($\sim -0.166$), indicating that the boundary
layer width scales
purely as the thermal diffusivity, i.e., $\delta_\mathrm{BL}\propto\sqrt{\kappa}$.
There is no strong dependence on viscosity within the range of \pr\ studied here.

We note that the highest Rayleigh number run had a boundary layer width of
about 3\% of the shell depth ($\sim 5$ Mm). Our thermal boundary layer lacks the radiative
processes at work in a star like the Sun, but its physical extent is confined to a
similarly small region of the convective domain.

Figure \ref{fig:dimKE}
showed that a free-fall regime that is independent of both viscosity and thermal
diffusion could be obtained for different Prandtl numbers. We find that in order to have a
bulk kinetic energy that is independent of both viscosity and thermal diffusion,
the thermal boundary layer thickness must depend on both the Rayleigh number and the
Prandtl number with the scaling law given in Equation \ref{eqn:wBLfit}.

For a fixed Prandtl number, this scaling relation suggests that
the boundary layer thickness will continually decrease as the Rayleigh number is increased.
In global simulations that employ explicit diffusivities,
the Prandtl number is therefore critical in maintaining
\refereeEdits{a boundary layer that is confined to a small region of the convective
domain without becoming vanishingly small as the Rayleigh number is increased.}

\refereeEdits{
Certainly, simulations with a conductive boundary layer cannot match every
feature of the Sun's thermal boundary layer, for the simple reason that the Sun's boundary
layer is regulated by radiative transfer instead of thermal conduction.
In a global simulation without radiative transfer, one hopes that the microphysics of the
cooling layer can be ignored and only the gross properties of the boundary layer
(thickness and entropy contrast) are important. When the boundary layer is conductive, the
entropy contrast and thickness are inherently linked. Thus, the best one could hope to do
is achieve a physically realistic thickness; this requires that the product
$H\left(\raf\pr\right)^{-1/6}$ take on the desired thickness, i.e., as the Rayleigh number
is increased, the Prandtl number must be decreased. This realistic thickness will
probably not, however, coincide with a convective power spectrum that possesses a realistic
inertial range.}

\section{Perspectives and Conclusions}\label{sec:perspective_conclusions}

The results presented here have interesting consequences for several
aspects of stellar/solar convection zone dynamics. Many of these results will depend
on rotation and magnetism, both of which were omitted in this study.

We find that simulations with a lower \pr\ number have faster flows and a broader
range of scales compared to high \pr\ models with the same \raf.
The higher \pr\ models have more viscous dissipation, resulting in slower
flows (equivalent to a lower Reynolds number).
This is consistent with the results found in \citet{omara2016}.

The higher \raf\ simulations obtain a free-fall state where diffusion no longer plays an
important
role in the interior, bulk global force balance. In this state the kinetic energy
becomes independent of both viscosity and thermal diffusion. Similar results
were found in \citet{featherstone2016}, although their study was restricted to $\pr=1$.

The boundary layer thickness scaling suggests that most simulations may be achieving the
correct driving scale with a
modest Rayleigh number of about $10^6$ and a Prandtl number of unity.
\refereeEdits{The obtained
scaling also suggests that in order to maintain a boundary layer whose physical extent
is confined to a small region of the convective domain in a
global simulation without radiative transfer, the
Prandtl number should be decreased as the Rayleigh number is increased.}

We did not find that the Prandtl number substantially alters the earlier observed
behavior in the spectral distribution of the velocity for \pr\ unity simulations
\citep[e.g.,][]{featherstone2016}.
As the \raf\ is increased, the convection develops smaller scale structures
and a corresponding increase in the high-wavenumber power. The high-wavenumber
power increases, but the low-wavenumber power decreases indicating the break down of
coherent large-scale structures. The spectral range $\ell \le 10$ appears to be the most
sensitive region of the power spectrum. This occurs for all Prandtl numbers studied here.

Our results indicate that care needs to be taken when interpreting convection
simulations and comparing the results to observations of real solar/stellar
systems. Simulations that do not access a high enough \raf, may overestimate the
low-wavenumber power that is accessible to helioseismology. Stellar
convection simulations must run with parameters that place it in the
high-Rayleigh-number regime in order to correctly capture the integrated kinetic energy and
the large scale motions of the flow.
We used rather modest levels of diffusion ($\kappa \le 4\times 10^{12}$ cm$^2$ s$^{-1}$)
to put our simulations in the high-Rayleigh-number regimes.

Most importantly, our simulations did not include rotation or magnetism.
\citet{featherstone2016b} looked at the effects of rotation, but restricted their
study to hydrodynamical, $\pr=1$ simulations.
It will be important to examine how our Prandtl number findings are modified by rotation
and magnetism before we can fully trust comparisons of power
spectra between observations and simulations.

\medskip
\noindent
This work was supported by NASA grants NNX09AB04G, NNX14AC05G, NNX11AJ36G, NNX17AM01G,
and NNX14AG05G.
Calkins was supported by the National Science Foundation award number EAR-1620649.
Featherstone and the development of \rayleigh\ were further supported by the Computational
Infrastructure for Geodynamics (CIG), which is supported by the National Science Foundation
award numbers NSF-0949446 and NSF-1550901. Resources supporting this work were provided by the NASA High-End Computing
(HEC) Program through the NASA Advanced Supercomputing (NAS) Division at Ames Research
Center.



\bibliography{Bibliography}

\begin{table*}[h]
{\centering
\begin{tabular*}{\textwidth}{cccccc@{\extracolsep{\fill}} cccccc}
\hline
\hline
\multicolumn{6}{c}{Input Parameters} & \multicolumn{6}{c}{Output Parameters} \\
\cline{1-6}
\cline{7-12}
\noalign{\vspace{2pt}}
$\kappa$ & $\nu$ &  \raf & \pr & $n_\mathrm{max}$ & $\ell_\mathrm{max}$ & KE &
$\widehat{\mathrm{KE}}$ & $f_\mathrm{conv}$ & $\delta_\mathrm{BL}$ & \re & $\re_\mathrm{peak}$ \\
\multicolumn{2}{l}{($10^{12}$ cm$^2$ s$^{-1}$)}&&&&&(10$^{38}$ erg) & & & (Mm) & & \\
\hline
\hline
10  & 1  & $6.88\times 10^4$ & 0.1 & 85 & 1023& 34.43 & 19064.7& 0.6620 & 15.57 & 248.3 & 342.1 \\
\hline
4  & 1  & $4.30\times 10^5$ & 0.25 & 85 & 1023& 42.59 & 23582.9& 0.8301 & 10.36 & 261.2 & 376.9 \\
8  & 2  & $5.37\times 10^4$ & 0.25 & 85 & 511 & 46.56 & 6446.4 & 0.7026 & 13.76 & 132.8 & 192.8 \\
12 & 3  & $1.59\times 10^4$ & 0.25 & 85 & 511 & 42.08 & 2588.8 & 0.5876 & 16.18 &  85.1 & 123.4 \\
16 & 4  & $6.72\times 10^3$ & 0.25 & 85 & 263 & 35.80 & 1238.4 & 0.5087 & 17.68 &  58.7 &  83.6 \\
24 & 6  & $1.99\times 10^3$ & 0.25 & 85 & 127 & 21.07 &  324.8 & 0.3251 & 20.78 &  30.8 &  44.9 \\
32 & 8  & $8.40\times 10^2$ & 0.25 & 85 & 127 & 9.36  &   81.6 & 0.1569 & 23.26 &  15.7 &  23.0 \\
\hline
4  & 2  & $2.15\times 10^5$ & 0.5 & 85 & 511 & 42.24 & 5847.2 & 0.8403 &  9.87 & 129.0 & 186.0 \\
6  & 3  & $6.37\times 10^4$ & 0.5 & 85 & 511 & 40.46 & 2489.6 & 0.7750 & 11.98 &  83.7 & 121.9 \\
8  & 4  & $2.68\times 10^4$ & 0.5 & 85 & 263 & 39.62 & 1371.2 & 0.7167 & 13.71 &  62.7 &  91.9 \\
12 & 6  & $7.96\times 10^3$ & 0.5 & 85 & 127 & 34.51 &  530.8 & 0.6031 & 16.18 &  38.5 &  56.5 \\
16 & 8  & $3.36\times 10^3$ & 0.5 & 85 & 127 & 24.84 &  214.8 & 0.4860 & 18.18 &  25.6 &  38.4 \\
24 & 12 & $9.96\times 10^2$ & 0.5 & 85 & 127 & 79.94 &   30.4 & 0.2314 & 22.21 &  10.4 &  16.5 \\
\hline
1  & 1  & $6.81\times 10^6$ & 1 & 85 &1023 & 36.22 &20057.7 & 0.9480 &  4.96 & 229.8 & 320.4 \\
2  & 2  & $8.53\times 10^5$ & 1 & 85 & 511 & 34.73 & 4807.6 & 0.9120 &  6.93 & 114.1 & 166.5 \\
4  & 4  & $1.06\times 10^5$ & 1 & 85 & 263 & 33.42 & 1156.7 & 0.8470 &  9.83 &  56.3 &  85.6 \\
6  & 6  & $3.16\times 10^4$ & 1 & 85 & 127 & 32.92 &  506.4 & 0.7910 & 12.18 &  37.5 &  58.2 \\
8  & 8  & $1.33\times 10^4$ & 1 & 85 & 127 & 31.62 &  273.5 & 0.7360 & 14.09 &  27.5 &  42.4 \\
12 & 12 & $3.94\times 10^3$ & 1 & 85 & 127 & 22.08 &   84.9 & 0.6230 & 16.03 &  15.4 &  24.0 \\
16 & 16 & $1.66\times 10^3$ & 1 & 42 & 127 & 12.02 &   26.0 & 0.4420 & 19.06 &   9.0 &  15.6 \\
24 & 24 & $4.93\times 10^2$ & 1 & 42 &  63 &  2.05 &    2.0 & 0.1220 & 23.77 &   2.5 &   4.5 \\
\hline
2  & 4  & $4.30\times 10^5$ & 2 & 85 & 511 & 27.22 & 941.9 & 0.9145 &  6.92 & 53.5 & 78.0 \\
4  & 8  & $5.37\times 10^4$ & 2 & 85 & 263 & 25.60 & 221.4 & 0.8539 & 10.18 & 26.6 & 40.0 \\
6  & 12 & $1.59\times 10^4$ & 2 & 85 & 127 & 24.37 &  93.7 & 0.8038 & 13.43 & 17.3 & 26.2 \\
8  & 16 & $6.72\times 10^3$ & 2 & 85 & 127 & 20.11 &  43.5 & 0.7398 & 15.22 & 12.1 & 18.7 \\
12 & 24 & $1.99\times 10^3$ & 2 & 85 & 127 & 10.04 &   9.6 & 0.5277 & 18.35 &  6.1 &  9.9 \\
16 & 32 & $8.40\times 10^2$ & 2 & 42 & 127 &  4.33 &   2.3 & 0.3393 & 20.65 &  3.1 &  5.2\\
\hline
1  & 4  & $1.72\times 10^6$ & 4 & 85 & 1023& 17.10 & 591.8 & 0.9490 & 5.02 & 45.3 & 67.1 \\
2  & 8  & $2.15\times 10^5$ & 4 & 85 & 511 & 20.26 & 175.2 & 0.9150 &  7.11 & 24.6 & 37.9 \\
4  & 16 & $2.68\times 10^4$ & 4 & 85 & 263 & 17.12 &  37.0 & 0.8641 & 11.62 & 11.7 & 18.5 \\
6  & 24 & $7.96\times 10^3$ & 4 & 85 & 127 & 11.88 &  11.4 & 0.7953 & 14.10 &  6.9 & 11.5 \\
8  & 32 & $3.36\times 10^3$ & 4 & 85 & 127 &  8.10 &   4.3 & 0.6808 & 15.86 &  4.2 &  7.1 \\
12 & 48 & $9.96\times 10^2$ & 4 & 85 & 127 &  3.66 &   0.8 & 0.4438 & 19.16 &  1.9 &  3.2 \\
16 & 64 & $4.29\times 10^2$ & 4 & 42 & 127 &  0.63 &   0.1 & 0.0969 & 24.08 &  0.6 &  1.0 \\
\end{tabular*}
\label{tab:params}
}
\caption{List of all the simulation parameters for each run. Each simulation used a
polytropic background state with an adiabatic index $\gamma=5/3$, three density scale heights
across the domain $N_\rho=3$, and a polytropic index of $n=1.5$. The inner and outer
radii of each simulation were $r_i = 5 \times 10^{10}$ cm and $r_o = 6.586\times 10^{10}$ cm.
The variable input parameters
are the thermal diffusivity $\kappa$, the kinematic viscosity $\nu$, the Rayleigh number,
the Prandtl number, the radial resolution, and the azimuthal resolution. The output parameters
are the dimensional kinetic energy KE, nondimensional kinetic energy $\widehat{\mathrm{KE}}$,
the fractional convective flux $f_\mathrm{conv}$, the dimensional thermal boundary layer
thickness $\delta_\mathrm{BL}$, the Reynolds number and the peak Reynolds number.}
\end{table*}

\end{document}